\documentclass[
amsmath,amssymb,
aps,
twocolumn,
showpacs,
prl,
]
{revtex4-1}

\usepackage{amsmath}    
\usepackage{graphicx}   
\usepackage{hyperref}   
\usepackage{dcolumn}    

\begin{document}

\title{Transverse Wobbling in $^{135}$Pr}

\author{J. T. Matta}
\affiliation{Physics Department, University of Notre Dame, Notre Dame, Indiana 46556, USA}
\author{U. Garg}
\affiliation{Physics Department, University of Notre Dame, Notre Dame, Indiana 46556, USA}
\author{W. Li}
\affiliation{Physics Department, University of Notre Dame, Notre Dame, Indiana 46556, USA}
\author{S. Frauendorf}
\affiliation{Physics Department, University of Notre Dame, Notre Dame, Indiana 46556, USA}
\author{A. D. Ayangeakaa}
\altaffiliation[Present address: ]{Physics Division, Argonne National Laboratory, Argonne, Illinois 60439, USA}
\affiliation{Physics Department, University of Notre Dame, Notre Dame, Indiana 46556, USA}
\author{D. Patel}
\affiliation{Physics Department, University of Notre Dame, Notre Dame, Indiana 46556, USA}
\author{K. W. Schlax}
\affiliation{Physics Department, University of Notre Dame, Notre Dame, Indiana 46556, USA}
\author{R. Palit}
\affiliation{Tata Institute of Fundamental Research, Mumbai 400 005, India}
\author{S. Saha}
\affiliation{Tata Institute of Fundamental Research, Mumbai 400 005, India}
\author{J. Sethi}
\affiliation{Tata Institute of Fundamental Research, Mumbai 400 005, India}
\author{T. Trivedi}
\altaffiliation[Present address: ]{Guru Ghasidas University, Bilaspur 495009, India}
\affiliation{Tata Institute of Fundamental Research, Mumbai 400 005, India}
\author{S. S. Ghugre}
\affiliation{UGC-DAE Consortium for Scientific Research, Kolkata 700 098, India}
\author{R. Raut}
\affiliation{UGC-DAE Consortium for Scientific Research, Kolkata 700 098, India}
\author{A. K. Sinha}
\affiliation{UGC-DAE Consortium for Scientific Research, Kolkata 700 098, India}
\author{R. V. F. Janssens}
\affiliation{Physics Division, Argonne National Laboratory, Argonne, Illinois 60439, USA}
\author{S. Zhu}
\affiliation{Physics Division, Argonne National Laboratory, Argonne, Illinois 60439, USA}
\author{M. P. Carpenter}
\affiliation{Physics Division, Argonne National Laboratory, Argonne, Illinois 60439, USA}
\author{T. Lauritsen}
\affiliation{Physics Division, Argonne National Laboratory, Argonne, Illinois 60439, USA}
\author{D. Seweryniak}
\affiliation{Physics Division, Argonne National Laboratory, Argonne, Illinois 60439, USA}
\author{C. J. Chiara}
\affiliation{Department of Chemistry and Biochemistry, University of Maryland, College Park, Maryland 20742, USA}
\affiliation{Physics Division, Argonne National Laboratory, Argonne, Illinois 60439, USA}
\author{F. G. Kondev}
\affiliation{Nuclear Engineering Division, Argonne National Laboratory, Argonne, Illinois 60439, USA}
\author{D. J. Hartley}
\affiliation{Department of Physics, U. S. Naval Academy, Annapolis, Maryland 21402, USA}
\author{C. M. Petrache}
\affiliation{Centre de Sciences Nucl\'{e}aires et Sciences de la Mati\`{e}re, Universit\'{e} Paris-Sud and CNRS/IN2P3, F-91405 Orsay, France}
\author{S. Mukhopadhyay}
\affiliation{Bhabha Atomic Research Centre, Mumbai 400 085, India}
\author{D. Vijaya Lakshmi}
\affiliation{Department of Nuclear Physics, Andhra University, Visakhapatnam 530 003, India}
\author{M. Kumar Raju}
\altaffiliation[Present address: ]{iThemba Labs, 7129 Somerset West, South Africa}
\affiliation{Department of Nuclear Physics, Andhra University, Visakhapatnam 530 003, India}
\author{P. V. Madhusudhana Rao}
\affiliation{Department of Nuclear Physics, Andhra University, Visakhapatnam 530 003, India}
\author{S. K. Tandel}
\affiliation{UM-DAE Centre for Excellence in Basic Sciences, Mumbai 400098, India}
\author{S. Ray}
\altaffiliation[Present address: ]{Amity University, Noida 201303, India}
\affiliation{Saha Institute of Nuclear Physics, Kolkata 700 064, India}
\author{F. D\"onau}
\altaffiliation{Deceased}
\affiliation{Institut f\"ur Strahlenphysik, Helmholtz-Zentrum Dresden-Rossendorf, 01314 Dresden, Germany}

\begin{abstract}
A pair of transverse wobbling bands has been observed in the nucleus $^{135}$Pr. The wobbling is characterized by $\Delta I$ =1, E2 transitions between the bands, and a decrease in the wobbling energy confirms its transverse nature. Additionally, a transition from transverse wobbling to a three-quasiparticle band comprised of strong magnetic dipole transitions is observed. These observations conform well to results from calculations with the Tilted Axis Cranking (TAC) model and the Quasiparticle Triaxial Rotor (QTR) Model. 
\end{abstract}

\maketitle
Deformed nuclei usually have an axial shape. The appearance of triaxial shapes at low to moderate spin has been predicted for a few limited regions of the nuclear chart, {\em e.g.} the  nuclei around $Z= 60$, $N=76$ and $Z=46$, $N=66$ \cite{Moller2006}. Calculations  predict that triaxial shapes become more common at high spin \cite{Werner1995}. There are two unique fingerprints of a triaxial nuclear shape: wobbling and chirality.  

Bohr and Mottelson had discussed wobbling of triaxial even-even nuclei many years ago \cite{NucStruc}. This mode represents
the quantized oscillations of the principal axes of an asymmetric top relative to the space-fixed angular momentum vector or, in the body fixed frame of reference, the oscillations of the angular momentum vector about the axis of the largest moment of inertia. The evidence for a triaxial shape is the inequality of the three moments of inertia, which is the prerequisite for the appearance of wobbling excitations. Clear evidence for wobbling in this purely collective form, which is seen in all asymmetric top  molecules, has not been found so far in the case of nuclei. Evidence for wobbling (collectively-enhanced E2 transitions between the the one - and zero-phonon rotational bands)
has been observed only in odd-A triaxial strongly deformed
(TSD) nuclei around $Z=72$, $N=94$ \cite{OdegardPRC.86.5866,Schonwasser20039,Bringel.EPJA.24.167,Amro2003197,Hartley.PRC.80.041304}.  However, in all these cases, the observed wobbling energy, $E_{wob}$ (defined later in the text), {\it decreases} with increasing angular momentum (see, for example, Ref. \cite{Hartley.PRC.80.041304}), in contrast with an {\it increase} expected for a purely collective wobbler and as evidenced in molecules. All these nuclei have an odd proton occupying an orbital with high intrinsic angular momentum, $j$, coupled to the triaxial  rotor, which considerably modifies the wobbling mode. Recently, Frauendorf and D\"onau \cite{Frauendorf.TW} have analyzed the modified mode, which they called ``transverse wobbling''. They identified the experimentally observed decrease in $E_{wob}$ as the hallmark of this mode, which they predicted to appear whenever a high-$j$ nucleon couples to a triaxial rotor core. It is important to verify this prediction and thus establish 
the presence of a triaxial shape. The odd-$Z$ nuclei with $A\sim{}130$ meet the condition: Triaxial shapes have been predicted  \cite{Moller2006} and the appearance of chirality, a complementary experimental evidence for triaxiality, has been established (see, for example, Refs. \cite{som,somprc}). 
 
In the scheme of transverse wobbling, the odd quasiproton, with predominantly particle nature, aligns its angular momentum vector $\vec j$ along  the short axis of the triaxial rotor. This arrangement is called  ``transverse'' because  the vector $\vec j$ is perpendicular to the axis with the largest moment of inertia (the medium axis) \cite{Frauendorf.TW}. Particle-like quasiparticles arising from the bottom of a deformed $j$-shell align their $\vec j$ vector with the short axis because this maximizes their overlap with the triaxial core, thus minimizing the energy of their attractive short-range interaction. This is the case for the odd h$_{11/2}$ proton in $^{135}$Pr and the nearby nuclei. 

Near the bandhead, the large $\vec j$ of the proton forces the total angular momentum vector to wobble about the {\it short} axis. Since the rotation is about a principal axis, signature is a good quantum number, being $\alpha(I)=mod(j,2)$ for the zero-phonon band and $\alpha(I)=mod(j,2)+1$ for the one-phonon band.  As angular momentum is added, rotation about the medium axis is energetically favored over that about the short axis, which has a smaller moment of inertia. There is a  critical angular momentum at which rotation about the short axis becomes unstable. At that point, the  rotational axis tilts  away from the short axis into the short-medium  principal plane. Consequently, transverse wobblers exhibit a decrease in the wobbling energy \cite{Frauendorf.TW}.

This Letter reports the first observation of  wobbling in the $A\sim{}130$ region. This is also the first observation of transverse wobbling at low deformation ($\epsilon{}\sim{}0.16$) based on the h$_{11/2}$ proton; the previously-observed cases involved the i$_{13/2}$ proton and significantly larger deformations ($\epsilon{}\sim{}0.40$) \cite{Frauendorf.TW}.
The partner of the yrast band of $^{135}$Pr that is interpreted as a transverse wobbler exhibits the expected characteristic of decreasing $E_{wob}$. Since wobbling is a strongly collective phenomenon, 
the $\Delta{}I=1$ interband transitions are expected to display primarily E2 character \cite{Frauendorf.TW}, which is confirmed by the $\gamma$-ray angular distribution and polarization measurements presented here.  Finally, theory predicts a three-quasiparticle dipole band that has magnetic nature, in accordance with the measurements. 

Two experiments were performed using the $^{123}$Sb($^{16}$O,4n)$^{135}$Pr reaction at a bombarding energy of 80 MeV. In the first one, carried out at the ATLAS facility at Argonne National Laboratory, the target was a 634 $\mu$g/cm$^2$-thick foil of isotopically enriched $^{123}$Sb, with a front layer of 15 $\mu$g/cm$^2$ Al. A total of $3.7\times10^{9}$ three- and higher-fold $\gamma$-ray coincidence events were collected using the Gammasphere array \cite{Lee1990c641}. The second experiment was carried out  at the TIFR-BARC Pelletron-LINAC facility at the Tata Institute of Fundamental Research, Mumbai, India. In this case, the target was 630 $\mu$g/cm$^2$ thick, sandwiched between a layer of 15 $\mu$g/cm$^2$ Al at the front and 20 mg/cm$^2$ Au at the back.  A total of  $4.5\times10^{8}$ two- and higher-fold $\gamma$-ray coincidence events were obtained with the Compton-suppressed clover array INGA \cite{Palit201290}. The data were analyzed using several software packages, including the RADWARE suite \cite{Radford1995297}, the BLUE libraries \cite{Cromaz2001519}, and the Multi pARameter time stamped based COincidence Search program (MARCOS) code for the INGA data.

\begin{figure}[h!]
  \begin{center}
\includegraphics[width=8.6cm]{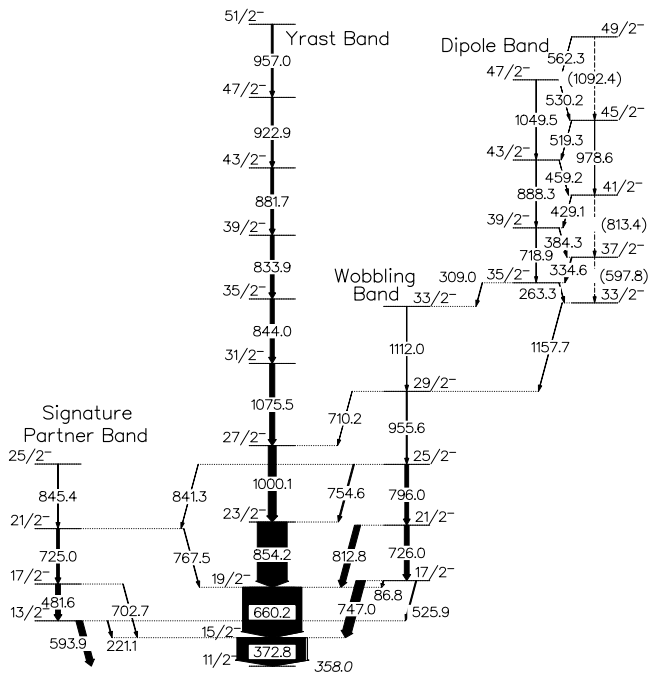}
\caption{\label{fig:partialScheme} Partial level scheme of $^{135}$Pr showing the previously known yrast band ($n_{\omega}$=0), the ``signature partner band'', the wobbling band ($n_{\omega}$=1), and the dipole band. The lowest level shown is an $\frac{11}{2}^{-}$ isomeric level with $E_x$ = 358 keV \cite{nndc}.}
\end{center}
\vspace{-0.7cm}
\end{figure}

A partial level scheme for $^{135}$Pr, based on detailed analysis of $\gamma$-$\gamma$-$\gamma$ coincidence relationships and highlighting the structures relevant to the focus of this Letter, is presented in Fig.~\ref{fig:partialScheme}; it builds on results previously reported for this nucleus \cite{SemkowPRC.34.523,PaulPRC.84.047302,epaul3,epaul2}. Spin and parity assignments for newly identified levels were made on the basis of DCO ratios, angular distributions, polarization measurements, and arguments from crossover $\gamma$-ray transitions. Details of the coincidence relationships and the individual angular distribution and asymmetry analyses, as well as the full level scheme, will be provided in forthcoming publications \cite{mattanext, mattanext2}.

The main features of the observed $^{135}$Pr level scheme are: the yrast band comprising a series of E2 transitions; a side band, also made up of E2 transitions (labeled the ``Wobbling Band'' in Fig.~\ref{fig:partialScheme}) and connected to the yrast band via $\Delta I=1$ transitions; a sequence of strong M1 transitions (the ``Dipole Band'') that builds on the wobbling band; and, a weak ``Signature Partner'' band.  The observed level scheme is in good agreement with previously-published results, except that the transitions belonging to the wobbling band were observed, but not correctly arranged in a band-like structure, in the low-statistics work presented in Ref. \cite{SemkowPRC.34.523}, and only the beginnings of the dipole band were observed in more recent unpublished data \cite{epaul3, epaul2}.

\begin{figure}[h!] 
\begin{center}
\includegraphics[width=8.6cm]{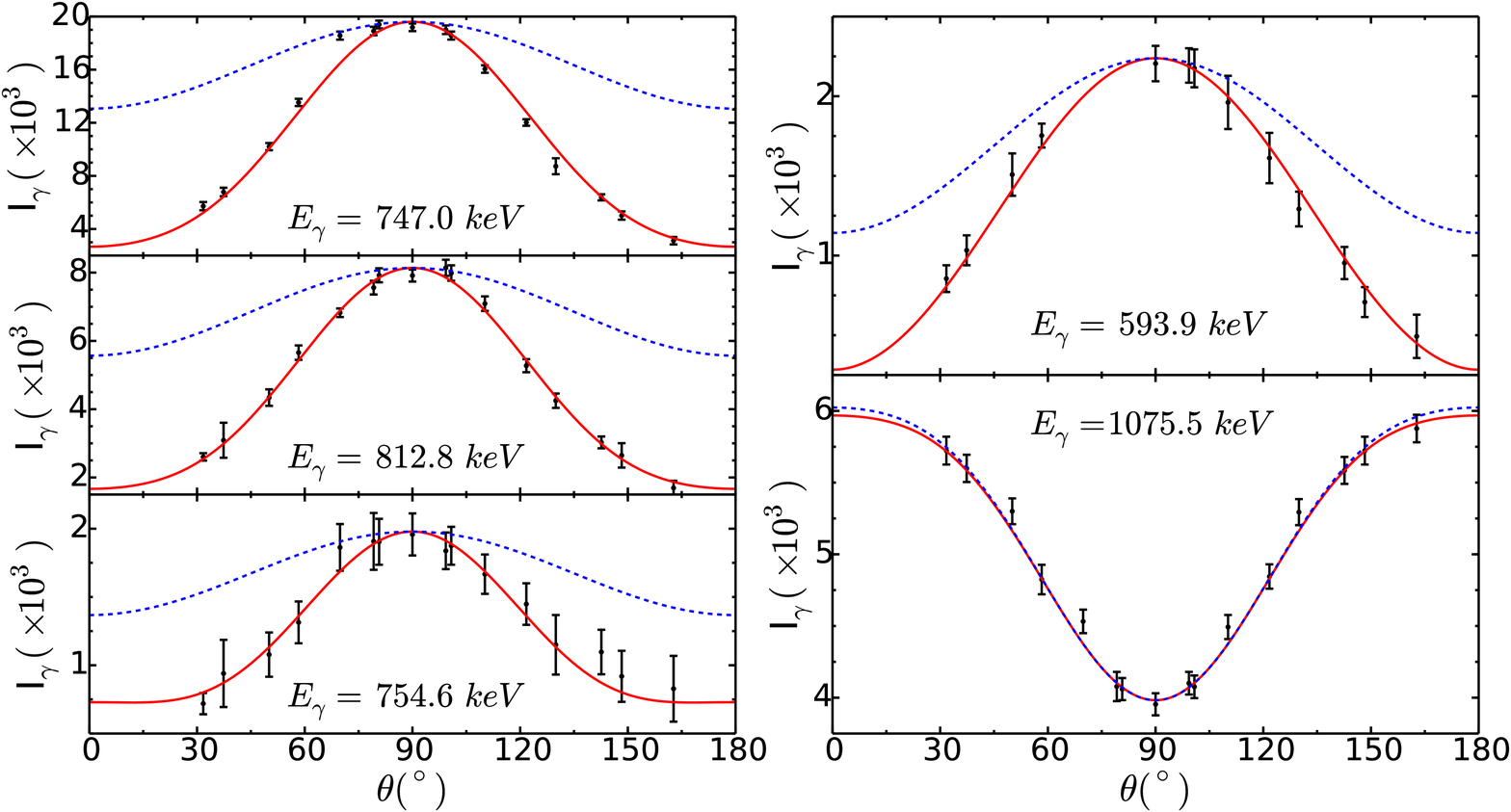}
\vspace{-0.7cm}
\caption{\label{fig:AngDists}(color online). (Left panels) Angular distributions for the first three transitions between the $n_{\omega}$=1 and $n_{\omega}$=0 bands, as also the best fits (solid red lines) from which the mixing ratios, $\delta$, presented in Table~\ref{table1}, were extracted (see text). The expected angular distributions for pure M1 transitions (dashed blue lines) are provided for comparison. Angular distribution data, and fits, are presented also for the 593.9-keV transition linking the signature-partner and yrast bands (see text) and for the 1075.5-keV pure E2 transition from the yrast band (right panels).}
\end{center}
\vspace{-0.5cm}
\end{figure}

For wobbling bands, the linking transitions are characterized by $\Delta I=1$, but are of E2 multipolarity, in contrast with the case of ``signature partner'' bands where the linking transitions are primarily M1. Indeed, the presence of linking transitions of the $\Delta I=1$, E2 type is a unique signature of wobbling bands \cite{OdegardPRC.86.5866}. To ascertain the nature of the transitions linking the wobbling band with the yrast band in $^{135}$Pr, angular distributions 
were analyzed using the data from Gammasphere, and the corresponding mixing ratios, $\delta$, extracted. The angular distributions were fitted with the function given in Ref.  \cite{Yamazaki1967}. The fits are presented in Fig.~\ref{fig:AngDists} and the resulting $\delta$ values are listed in Table \ref{table1}. 

\begin{figure}[h!]
  \begin{center}
\includegraphics[width=8.6cm]{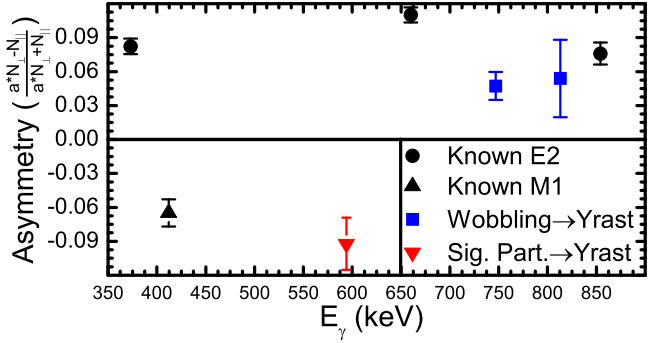}
\vspace{-0.7cm}
\caption{\label{fig:Asyms} (color online). The asymmetries (filled squares) for the first two transitions (747.0- and 812.8-keV) from the $n_{\omega}$=1 band to the $n_{\omega}$=0 band, as extracted from polarization data. The filled circles represent known E2 transitions from the yrast band and the black triangle a known 412.3-keV M1 transition from a level sequence in $^{135}$Pr not displayed in Fig.~\ref{fig:partialScheme}. Also shown is the measured polarization asymmetry for the 593.9-keV transition linking the signature-partner and yrast bands (see text). }
\end{center}
\vspace{-0.7cm}
\end{figure}

The large mixing ratios correspond to high E2 admixtures---up to 85\% for the highest transition for which angular distribution data were reliably obtained. To conclusively establish the predominantly electric nature of the linking transitions, polarization asymmetries were determined for the relevant transitions from data obtained with the INGA array.
In the two cases where the data had sufficient statistics to reliably extract the asymmetries (see Ref. \cite{Starosta199916} for details),  the asymmetry parameter is $>$0, clearly identifying these transitions as predominantly electric in nature. The measured asymmetry parameters are presented in Fig.~\ref{fig:Asyms}. We note that, in contrast, both the angular distribution and polarization asymmetry for the 593.9-keV, $\frac{13}{2}^{-}\rightarrow \frac{11}{2}^{-}$ transition from the signature-partner band to the yrast band establish its primarily M1 character. All other transitions interlinking the signature-partner and yrast bands are too weak for extraction of full angular distributions or determination of polarization asymmetries; however, the extracted DCO ratio for the 707.2-keV, $\frac{17}{2}^{-}\rightarrow \frac{15}{2}^{-}$ transition leads to a pure dipole assignment for this transition as well. 

The $\Delta I$=1, E2 character of the transitions linking the main and wobbler bands in $^{135}$Pr clearly establishes these bands as a wobbler pair corresponding to $n_{\omega}$=0 and $n_{\omega}$=1, respectively.

\begin{table*}
\begin{center}
\caption{Mixing ratios, $\delta$, E2 fractions, and the experimental and theoretical transition probability ratios for transitions from the $n_{\omega}$=1 to $n_{\omega}$=0 wobbling bands 
in $^{135}$Pr. The in-band transitions were assumed to be of pure E2 character in calculations of the probability ratios.
The mixing ratio of the $\frac{25}{2}^-\rightarrow \frac{23}{2}^-$ transition has been taken as a lower limit when deriving the probability ratios for the $\frac{29}{2}^-\rightarrow \frac{27}{2}^-$ transition. Shown at the bottom is the measured mixing ratio for the lowest signature partner to yrast transition.}
\label{table1}
\begin{tabular}{|c|c|c|c|c|c|c|c|c|c|}
\hline
Initial $I^\pi{}$ & Final $I^\pi{}$ &$E_{\gamma}$ &$\delta$&Asymmetry & E2 Fraction & \multicolumn{2}{c|}{ $\frac{B(M1_{out})}{B(E2_{in})}(\frac{\mu_N^2}{e^2b^2})$} & \multicolumn{2}{c|}{$\frac{B(E2_{out})}{B(E2_{in})}$} \\
$n_{\omega}=1$ & $n_{\omega}=0$ & (keV) & & &  (\%) & Experiment & QTR & Experiment & QTR \\
\hline
$\frac{17}{2}^-$ & $\frac{15}{2}^-$ & 747.0 & $-1.24\pm0.13$ & $0.047\pm0.012$ & $60.6\pm5.1$ & -- & $0.213$ & -- & 0.908 \\
$\frac{21}{2}^-$ & $\frac{19}{2}^-$ & 812.8 & $-1.54\pm0.09$ & $0.054\pm0.034$ & $70.3\pm2.4$ &  $0.164\pm0.014$ & $0.107$ &  $0.843\pm0.032$ & 0.488 \\
$\frac{25}{2}^-$ & $\frac{23}{2}^-$ & 754.6 & $-2.38\pm0.37$ & -- & $85.0\pm4.0$ &   $0.035\pm0.009$ & $0.070$ &  $0.500\pm0.025$ & 0.290 \\
$\frac{29}{2}^-$ & $\frac{27}{2}^-$ & 710.2 & -- & -- & -- & $\leq0.016\pm0.004$ & $0.056$ & $\geq 0.261\pm0.014$ & 0.191 \\
\hline
$\frac{13}{2}^-$ & $\frac{11}{2}^-$ & 593.9 & $-0.16\pm0.04$ & -- & $2.5\pm1.2$ & -- & -- & -- & -- \\
\hline
\end{tabular}
\end{center}
\end{table*}

\begin{figure}[h!]
  \begin{center}
\includegraphics[width=8.6cm]{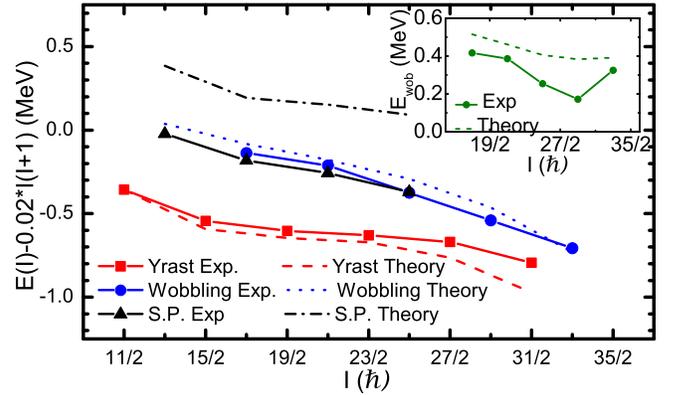}
\vspace{-0.7cm}
\caption{\label{fig:EWobVsI} (color online). Experimental and QTR level energies minus a rotor contribution for the yrast, wobbling, and signature partner bands. Inset: Wobbling energies (filled circles) associated with the wobbler-band pair observed in the nucleus $^{135}$Pr plotted as a function of spin $I$. Predicted values from the QTR model are also shown (see text). The lines through the experimental points are drawn to guide the eye.}
\end{center}
\vspace{-0.7cm}
\end{figure}

The wobbling energies, $E_{wob}$, defined as:
\begin{eqnarray}
E_{wob} (I) &=& E (I,n_{\omega}=1) - [ E (I-1,n_{\omega}=0) \nonumber \\
            && + E (I+1,n_{\omega}=0) ]/2,
\end{eqnarray}
were calculated from the level energies and are presented in the inset of Fig.~\ref{fig:EWobVsI} as a function of the spin, $I$. The wobbling energy decreases with angular momentum--this is the hallmark of transverse wobbling. The combination of the nature of the interlinking transitions {\em and} the $E_{wob} (I)$ vs. $I$ behavior  firmly identifies the observed level structure in $^{135}$Pr as arising from a transverse wobbler.

\begin{figure}[h!]
\begin{center}
\includegraphics[width=8.6cm]{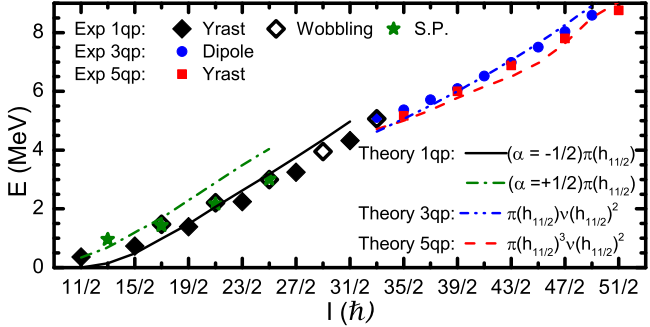}
\vspace{-0.7cm}
\caption{\label{fig:EvsJ} (color online). Level energies for the three bands featured in Fig.~\ref{fig:partialScheme}; the filled and open diamonds represent the wobbling band pair and the filled squares represent the three-quasiparticle dipole band. The results from the TAC model are presented as lines.}
\end{center}
\vspace{-0.7cm}
\end{figure}

TAC mean-field calculations  \cite{Frauendorf2000115} were carried out  for the one-quasiproton yrast band. Using the pairing gaps $\Delta_p=1.1$ MeV and $\Delta_n=1.0$ MeV, we obtained  equilibrium deformation parameters  $\epsilon=0.16$ and $\gamma=26^\circ$, which were kept constant. Additional TAC calculations were carried out for the [$\pi$h$_{11/2}$, $\nu$h$_{11/2}^2$] three-quasiparticle, and the [$\pi$h$_{11/2}^3$, $\nu$h$_{11/2}^2$] five-quasiparticle configurations. Using  the pairing gaps $\Delta_n=0$ and $\Delta_p=0.8$ MeV resulted in approximately constant deformation parameters of  $\epsilon=0.20$, $\gamma=28^\circ$.  The short axis was found to be the stable axis of rotation for the one-quasiproton yrast band. The 
same holds for the five-quasiparticle configuration, which we interpret, in agreement with Ref. \cite{PaulPRC.84.047302}, as the continuation of the $\Delta I=2$  yrast sequence after 
the combined alignment of a quasiproton and quasineutron pair. In this configuration, the two quasineutrons occupy h$_{11/2}$ orbitals 
with opposite angular momentum projection on the long axis which results in no
preference of one of the principal axes. In the case of the the three quasiparticle configuration, the two quasineutrons occupy h$_{11/2}$ orbitals 
with the same angular momentum projection on the long axis. The combination of the neutron alignment with the long axis and the proton alignment with the short axis results in an angle of 
67$^\circ$  between the long and 
rotational axes. Accordingly, the corresponding $\Delta I=1$ band has dipole character.  As seen in 
Fig.~\ref{fig:EvsJ}, the TAC calculations reproduce the experimental energies fairly well. 
The calculated intraband ratios $B(M1, I\rightarrow I-1)/B(E2,I\rightarrow I-2)\approx 3.0/0.55=5.4~\mu_N^2/(eb)^2$ indicate dominance of  magnetic rotation. 
Experimental intraband ratios varied from $3.0\pm0.4$ $\mu_N^2/(eb)^2$ to $18.8\pm0.9$ $\mu_N^2/(eb)^2$.

Fig. \ref{fig:EWobVsI}  compares the 
experimental energies with those calculated by means of a modification of the Quasiparticle+Triaxial Rotor (QTR) model based on the 
Quasiparticle Core Coupling model of Ref. \cite{DoenauQCC}. 
The triaxial rotor is parametrized by three angular momentum-dependent moments of inertia
${\cal J}_i=\Theta_i(1+cI)$, where $i=m,s,l$ denotes the medium, short, long axes, respectively. The parameters ${\cal J}_m ,{\cal J}_s, {\cal J}_l$ = 7.4, 5.6, 1.8 $\hbar^2$/MeV, and c=0.116  
 were determined by adjusting the QTR energies to the experimental energies of the zero- and one-phonon bands. 
 The corresponding moments calculated by the TAC model, ${\cal J}_m ,{\cal J}_s, {\cal J}_l$ = 19, 8, 3 $\hbar^2$/MeV, respectively, result in the moment ratios ${\cal J}_m /{\cal J}_s /{\cal J}_l$ = 1/0.42/0.16, which 
 lead to too early a collapse of the transverse wobbling 
 regime. This is avoided by the fitted ratios ${\cal J}_m /{\cal J}_s /{\cal J}_l$ = 1/0.75/0.24, reflecting the fact that the wobbling mode is stabilized by the larger value for the ratio ${\cal J}_s /{\cal J}_m$ (see Ref. \cite{Frauendorf.TW}
 for details). 
 The QTR calculations for the zero-  and  one-phonon  wobbling states are in fair agreement with the data. $E_{wob}$ decreases first, as is characteristic for the transverse wobbler, but turns upward after reaching a minimum at $I^\pi=\frac{29}{2}^-$. 
The reason is that the Coriolis force detaches the $\vec j$ vector of the h$_{11/2}$ quasiproton from the short axis and aligns it with the medium axis.
The experimental wobbling energies  show a more pronounced minimum, which largely reflects the onset of the transition to the 
five-quasiparticle configuration in the high-spin yrast structure. 

As seen in Table \ref{table1}, the QTR model predicts a strong, non-stretched E2 component which dominates the M1 part in the mixed transitions de-exciting the one-phonon wobbling band. However, the calculations underestimate the strong $B(E2_{out})$ transition probabilities somewhat, and overestimate the weak $B(M1_{out})$ counterparts. 
The QTR calculations also predict a second signature $\alpha=1/2$ band, which is interpreted as the one-quasiproton signature partner of the $\alpha=-1/2$ yrast  band.  
The very small $B(E2_{out})/B(E2_{in})<0.01$ and $B(M1_{out}/B(E2_{in}))<0.02\mu_N^2/e^2b^2$ values are characteristic for transitions $I\rightarrow I-1$ between signature partners close to decoupling; for comparison, the estimated experimental values for these ratios for the $\frac{17}{2}^-\rightarrow \frac{15}{2}^-$ transition are 0.0002 and 0.004, respectively. The QTR calculations
predict the signature partner band $\sim$500-keV too high (Fig. \ref{fig:EWobVsI}); the TAC calculation, however, gives about the right excitation energy (Fig. \ref{fig:EvsJ}).

In summary, we have investigated the phenomenon of transverse wobbling in the A$\sim$130 region. A wobbler partner band has been identified in the nucleus $^{135}$Pr, the first observation of wobbling in a mass region other than A$\sim$160.   The nature of wobbler bands is confirmed by verifying the $\Delta I$=1, E2 character of the interband transitions via angular distribution and polarization measurements. The transverse nature of wobbling is evidenced by the characteristic decrease in  the wobbling energy, $E_{wob}$. 
In addition, a second band, with signature opposite to the yrast band, was identified with the characteristics of a signature partner.
 The appearance  of a collective  wobbling excitation, in addition to the signature partner
quasiparticle excitation, is clear evidence for deviation from axiality. The wobbling structure mutates into a three-quasiparticle band of the magnetic rotation type. All these observations are in good agreement with calculations in the framework of TAC and QTR models. A systematic search for similar wobbling band structures in the nearby nuclei is imperative.

We thank Dr. Michael Albers for helpful discussions about the angular distribution analysis procedures. This work has been supported in part by the U. S. National Science Foundation (Grants No. PHY-1068192 (UND) and No. PHY-1203100 (USNA)); by the APS-IUSSTF Physics Student \& Post-doc Visitation Program; by the U. S. Department of Energy, Office of Science, Office of Nuclear Physics, under Contracts No. DE-AC02-06CH11357 (ANL), No. DE-FG02-95ER40934 (UND) and DE-FG02-94ER40834 (UMCP); and, by the Department of Science and Technology, Government of India (Grants No. IR/S2/PF-03/2003-II and No. IR/S2/PF-03/2003-III). This research used resources of ANL's ATLAS facility, which is a DOE Office of Science User Facility.

\bibliographystyle{apsrev}

\bibliography{135PrPRLPaper}

\end{document}